\documentclass[prl,floatfix,twocolumn,showpacs]{revtex4}
\usepackage{amsmath}
\usepackage{amsfonts}
\usepackage{mathrsfs}
\usepackage{epsfig}
\usepackage{graphicx}
\usepackage{dcolumn}
\usepackage{amsmath}

\def\be{\begin{equation}}
\def\ee{\end{equation}}

\begin{document}

%\title{Activity-enhanced response of excitable media to sub-threshold
%stimulation}
\title{Response to sub-threshold stimulus is enhanced by spatially
heterogeneous activity}
%\title{Spatially heterogeneous activity enhances response to sub-threshold
%stimulation in excitable media}
%\title{Control of global chaos in excitable media through the application of a global sub-threshold signal}

\author{S. Sridhar and Sitabhra Sinha}
\affiliation{%
The Institute of Mathematical Sciences, CIT Campus, Taramani,
Chennai 600113, India.}
 \date{\today}
\begin{abstract}
Sub-threshold stimuli cannot initiate excitations in active media, but
surprisingly as we show in this paper, they can alter the
time-evolution of spatially heterogeneous activity by modifying the
recovery dynamics. This results in significant reduction of waveback
velocity which may lead to spatial coherence, terminating all activity
in the medium including spatiotemporal chaos. We analytically derive
model-independent conditions for which such behavior can be observed.
%Sub-threshold stimuli, by definition, has no significant effect on the
%excitation behavior of active media. In this paper, we show that these
%stimuli can have surprising consequences on the
%time-evolution of spatially heterogeneous activity, as a result of
%their effect on the recovery dynamics of such media.
%The reduction of waveback velocity induced by sub-threshold stimulation 
%leads to spatial coherence that can terminate all activity in the
%medium, including spatiotemporal chaos. We analytically
%derive a relation between the stimulation parameters for which such
%behavior can be observed. 
%We further develop a simple
%mechanism to explain the conditions leading to the termination of all
%activity in the medium and determine the  minimum duration of the weak
%stimulus required to achieve this.
\end{abstract}

 \maketitle
  
 %%%%
 \newpage
 %%%%

Occurrence of spatiotemporal patterns is a generic feature of a wide variety of 
natural systems that are excitable~\cite{Cross93}. Examples of such patterns include 
%Occurrence of spatiotemporal patterns, such as 
traveling waves in the
Belousov-Zhabotinsky reaction~\cite{Zaikin70} and propagating fronts
of surface catalytic oxidation~\cite{Jakubith90}. 
%is a generic feature 
%of a wide variety of natural systems sharing the property of being
%excitable~\cite{Cross93}. 
Excitability in these systems is characterised 
by a threshold. A supra-threshold stimulus causes a transition from a
quiescent to an active state, generating an action potential (AP).  
Following this excitation, the system slowly returns to rest.
%following the action potential (AP) resulting 
During this recovery period the system
is either fully or partially insensitive to another identical stimulus.
%The distinctive patterns observed in excitable media 
The dynamical consequences of these properties result in distinctive
patterns, such as rotating spiral waves, that may in turn lead to
spatiotemporal chaos~\cite{Winfree84,Keener98,Agladze94}.
While the threshold is a key parameter governing excitable systems,
the demonstration of stochastic resonance (SR)~\cite{Jung95} and coherence
resonance (CR)~\cite{Kurths97} in such media suggest that weak sub-threshold
signals could also have a significant
effect on their dynamics~\cite{Alonso01,Muratov07}.
%While it would appear that a
%sub-threshold signal will have negligible effect on the evolution of
%such patterns, it is known that excitable media exhibit
SR-like response resulting from chaotic dynamics in
simple systems~\cite{Sinha98} raises the intriguing possibility
that spatially heterogeneous activity may enhance the
response of an excitable medium to sub-threshold signals.
%it is possible that sub-threshold stimulation of an excitable medium
%with spatially heterogeneous activity may have surprising
%consequences.

Understanding how weak signals affect spiral waves and other
spatiotemporal activity in excitable media is especially important
because such patterns have critical functional consequences for
vital biological systems.
%special importance owing to the critical functional
%consequences of such patterns in certain biological systems.
For example, rotating vortices in cardiac
tissue that can lead to spiral chaos underlie many
arrhythmias, i.e., life-threatening disturbances in the natural rhythm
of the heart~\cite{Gray98,Witkowski98}.
%It is thus crucial to understand how the strength of the stimulus
%affects the evolution of these dynamical patterns.
%Thus, using low-amplitude stimulation for
Thus, controlling irregular activity in excitable
media is not only a problem
of fundamental interest in the physics of nonlinear dynamical
systems but also
has potential clinical significance~\cite{Garfinkel92,Alonso03}.
%in efficiently terminating
%cardiac arrhythmias~\cite{Garfinkel92,Alonso03}.
%irregular spatiotemporal activity associated with
Existing methods of spatiotemporal chaos control in excitable systems
are almost exclusively
dependent on using supra-threshold signals, either through a local
high-frequency source~\cite{Zhang05,Pumir10} or using
a spatially extended array~\cite{Sinha01}. Controlling spatial patterns with
sub-threshold stimulation would not only utilize new physical principles, 
but also avoid many of the drawbacks
in previously proposed schemes.

In this paper we show that sub-threshold stimulation, while having no
significant effect on a quiescent medium, can induce
a remarkable degree of coherence when applied on a system with spatially
heterogeneous activity. 
Synchronizing the state of activation of all excited regions ensures
that they return to rest almost simultaneously, in the process
completely terminating activity in the medium. Thus, control of
spatially extended chaos is achieved efficiently using a very
low-amplitude signal. We explain the mechanism of this enhanced coherence 
in terms of the role played by sub-threshold stimulus in increasing the
recovery period of the medium. It significantly reduces the
propagation velocity of the recovery front, thereby increasing the
extent of the inexcitable region in the medium.
We present a semi-analytical
derivation of the relation between strength and duration of the
globally applied sub-threshold signal necessary for complete
elimination of spatially heterogeneous activity in excitable media.

A generic model for describing the spatiotemporal dynamics of several
biological excitable systems has the form:
\begin{equation} \label{eq1}
\frac{\partial V}{\partial t}    = \frac{- I_{ion}(V,g_i) + I(t)}{C_m} +
D{\nabla}^2 V,
\end{equation}
where $V$ (mV) is the potential difference across a cellular membrane,
$C_m (= 1 \mu {\rm F} {\rm cm}^{-2})$ is the transmembrane
capacitance, $D$ is the diffusion constant
(~$=0.001 {\rm cm}^2 {s}^{-1}$ for the results reported in the paper), 
$I_{ion} (\mu{\rm A cm}^{-2})$ is the total current density 
through ion channels on the cellular membrane,
and $g_i$ describes the dynamics
of gating variables of different ion channels. 
The spatially uniform external signal, applied at all points of the simulation
domain, is represented by the time-dependent current density,
$I (\mu{\rm A cm}^{-2})$.
The specific functional form for $I_{ion}$ varies for different
biological systems.
%and depends on system-specific details.
For the results reported here, we have used the Luo-Rudy I
(LR1) model that describes the ionic currents in a
ventricular cell~\cite{LR1}. 
%of the guinea pig heart
%{\bf Give $g_{si}$ and other parameter values which differ from
%published LR values.}
For all our simulations, the maximum $K^+$ channel conductance $G_K$
has been increased to 0.705 mS cm$^{-2}$ to reduce the duration of the
action potential (APD)~\cite{tenTusscher03}.
To study the effect of sub-threshold stimulus on a stable spiral
and on spatiotemporal chaos, we have used the 
maximum $Ca^{2+}$ channel conductance $G_{si} = 0.04$ and 
0.05 mS cm$^{-2}$, respectively.
We have explicitly verified the model-independence of our results by 
observing similar effects in other realistic channel-based descriptions of 
the ionic current, such as the TNNP model~\cite{tenTusscher04,Sridhar10}.

%One area of active interest in the study of excitable medium is the
%effect of stochastic noise on a sub-excitable medium that by definition cannot support
%patterns. Another way to look at the problem is to study the resonance of the external noise with 
%a sub-threshold signal. Formation and breakup of spiral waves in an excitable
%medium due to the effect of such stochastic resonance was first observed in a
%model of cellular automata~\cite{Jung95}. The externally applied noise leads to 
%the formation of spiral waves and spiral breakup for external periodic signal
%strengths below the pattern forming threshold. Pulsating spots of activity~\cite{Hempel99} and target
%patterns~\cite{Muratov07} have also been observed in sub-excitable regimes.
%Experimentally, travelling waves and target wave patterns supported by 
%external noise have been observed in sub-excitable chemical systems~\cite{Showalter98,Alonso01}.
%It has also been observed that even in the absence of external periodic signal, 
%there is a resonance like phenomenon observed when the external noise interacts with
%the non-linear activity in the medium termed coherent resonance ~\cite{Kurths97,Carrillo04}.

We consider in turn the response of a single cell, a 1-dimensional
cable and a 2-dimensional sheet of excitable units to a
sub-threshold current $I$.
The spatially extended systems are discretized on a grid of size $L$
(for 1-D) and $L \times L$ (for 2-D). For most results reported here
$L = 400$, although we have used $L$ upto $1200$. The space step used
for all simulations is  $\delta x = 0.0225$ cm, while the time-step
$\delta t = 0.05$ ms (for 1-D) or 0.01 ms (for 2-D). 
The equations are solved using a forward Euler scheme with a 
standard 3-point (for 1-D) or 5-point (for 2-D) stencil for the
Laplacian describing the spatial coupling between the units.
%The diffusion constant $D=0.001 cm/sec^2$ and transmembrane
%capacitance $C_m=1\mu F/cm^{-2}$. 
No-flux boundary conditions are
implemented at the edges. The external current is applied globally,
i.e., $I (t)=I$ in Eq.~(\ref{eq1}) at all
points in the system for the duration of stimulation, $\tau$. 
A stimulus $\{ I, \tau \}$ is sub-threshold if it does not generate an action
potential when applied on a quiescent medium.
%To consider the effect of sub-threshold stimuli on
%spatiotemporally activity in excitable media, we apply them on
%states exhibiting either a single spiral or spatially extended chaos.

\begin{figure}
\centerline{
\includegraphics[width=1.0\linewidth,clip]{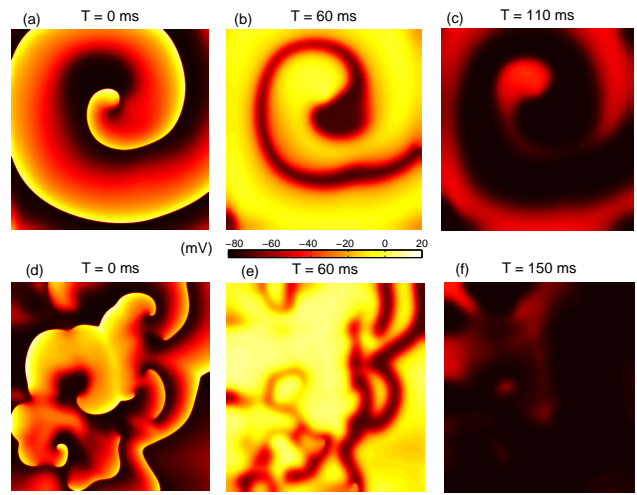}
}
\caption[]{
Pseudocolor plots of transmembrane potential $V$ for the two-dimensional
LR1 model ($L = 400$) showing the elimination of all activity on
applying a
sub-threshold current. The current $I$ is switched on at $T = 0$
for a duration $\tau= 60$ ms 
on (a-c) a single spiral, with $I$
=$1.6 \mu$A cm$^{-2}$, and (d-f) a spatiotemporally chaotic state, 
with $I$ =$1.8 \mu$A cm$^{-2}$.
By $T=150ms$, excitation has been effectively terminated throughout
the simulation domain.
}
\label{Figure1}
\end{figure}

\begin{figure}
\centerline{
\includegraphics[width=1.0\linewidth,clip]{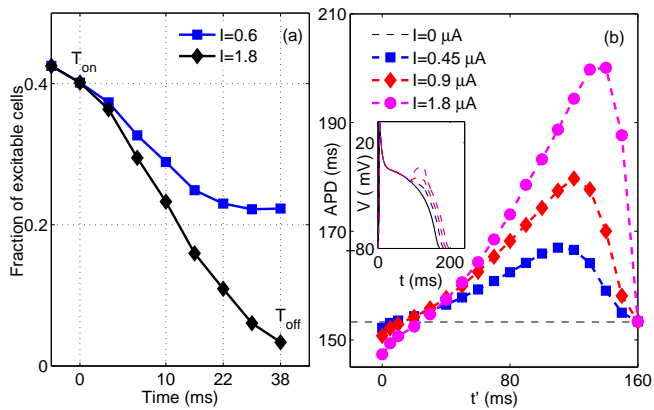}
}
\caption{(a) The fraction of cells that can be potentially excited
(i.e., with $V \leq -60mV$) decreases with time
during stimulation by sub-threshold current $I$ in a
two-dimensional system with spatiotemporal chaos. The total duration
of the external signal is $\tau = T_{on}-T_{off} = 38$ ms.
%The times at which the external stimulation is initiated and terminated
%are indicated by $T_{on}$ and $T_{off}$ respectively.
Results shown
correspond to failed (squares) and successful termination
(diamonds) of activity in the medium. 
(b) Effect of sub-threshold current on APD of a single cell. 
The current $I$ is varied keeping the duration $\tau$ fixed ($= 38$ ms). 
In all cases, the APD is shown as a function of the time interval
$t^{\prime}$ between the initiation of AP and
%between the time instant at which the AP is initiated and
$T_{on}$.
%to the initiation of the AP at which the external signal is switched on. 
The inset shows the
corresponding effect
%the sub-threshold stimulus
on the AP profile.
%In both cases, $T_1$ and $T_2$ indicate the times at which the
%external stimulation is switched on and off, respectively.
}
\label{Figure2}
\end{figure}

Fig.~\ref{Figure1} shows that when sub-threshold stimulation is
applied to an excitable medium with spatially heterogeneous activity,
viz., either a single spiral wave (a-c) or
spatiotemporal chaos (d-f), there is a striking change on the subsequent
dynamics of the system.
%We first apply sub-threshold signal to a 2-dimensional excitable medium having
%spatially heterogeneous activity. 
Within a short duration (comparable to the APD) there is
complete suppression of all activity in the medium, although in
absence of this intervention, the existing dynamical state would continue
to persist for an extremely long time.
This result is surprising as the weak sub-threshold signal 
appears to be incapable of 
significantly altering the dynamics of an excitable system.
To understand this apparent paradox, we first note that the
sub-threshold stimulation rapidly decreases the number of cells that
can be excited by existing activity in the medium (Fig.~\ref{Figure2},
a).
%that the sub-threshold stimulation results in a rapid decrease in the number
%of excitable cells in the medium, i.e., cells which can give rise to
%an action potential immediately on stimulation, throughout the duration 
%that the stimulus is maintained. 
Indeed, global suppression
of activity results when, by the end of the stimulation, the number of
cells susceptible to excitation is insufficient to sustain
the activity.
This decrease in their number is because cells tend
to remain in the recovering state for a longer period 
in the presence of a sub-threshold stimulus.
We can see this clearly in the response of a single {\em
excited} cell to a subsequent sub-threshold current $I$ applied for a
fixed duration $\tau$ (Fig.~\ref{Figure2},b).
%an effect that can be seen clearly for a single excitable cell.
%We want to say that the only effect of the sub-threshold stimulus is
%on the recovering phase of the excitable cell
%In Fig.~\ref{Figure2}~(b), we observe 
%the response of an {\em excited} cell to a subsequent
%sub-threshold current $I$ applied for a fixed duration $\tau$ after time
%$t^{\prime}$ from the initiation of the AP.
Increasing $I$ significantly alters the recovery 
period resulting in a change in the APD.
%Increasing $\tau$, keeping $I$ constant, has a similar effect (Figure
%not shown).
The time $t^{\prime}$ (measured from the initiation of the AP)
at which the sub-threshold stimulation begins also
affects the response of the cell to the signal. 
%For example, for
%$t^{\prime} > 20$ ms 
%there is a significant increase in the APD.
These results clearly indicate that the dominant effect of a sub-threshold
stimulus is to increase the time period that a cell spends in
recovering from prior excitation.
%we look at the response of a single
%excitable cell, in terms of the profile of its action potential (AP), 
%to a sub-threshold stimuli (Fig.~\ref{Figure3}). First, the cell is
%stimulated with a supra-threshold signal to generate an AP.
%Next, a sub-threshold signal is applied after different
%intervals, i.e., at different phases of the AP.
%The applied signal is characterised by its amplitude $I$ and
%duration $\tau$, increasing either of which results in an enhanced
%alteration of the repolarization phase of the AP.
%This is reflected in the change of the action potential duration
%(APD), as a function of $I$, $\tau$ and the time at which it is
%applied after the initiation of the AP. When the sub-threshold signal
%is applied very close to AP generation, the APD is slightly decreased.
%However, if the signal is applied at a later time ($> 20$ ms after AP
%initiation) there is a significant increase in the APD. In other words,
%the cell remains in the recovering state for a longer period in the
%presence of the sub-threshold stimulation.
\begin{figure}[t]
\centerline{
\includegraphics[width=1.0\linewidth,clip]{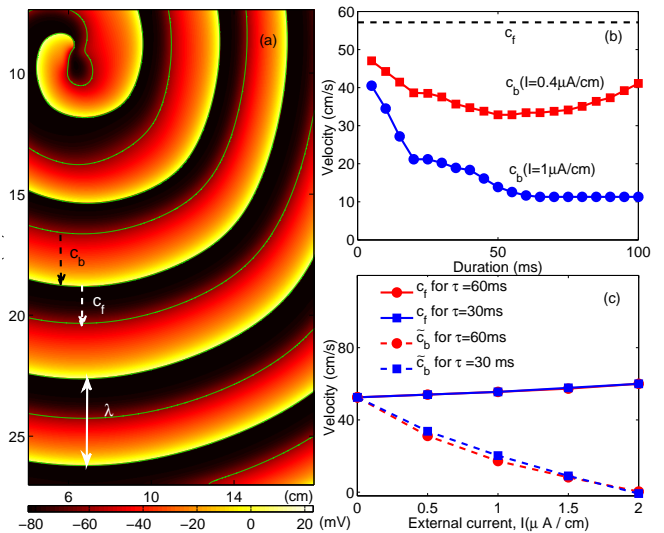}
}
\caption[]{
%(a) A magnified section of a 
(a) Pseudocolor plot of a spiral wave 
indicating the wavelength $\lambda$, the wavefront
velocity $c_f$ and the instantaneous waveback velocity $c_b (= c_f)$ in the absence
of external stimulation. 
%The total system size is
%$18$cm $\times$ $27$cm and the mean action potential duration, $\tau_{AP} =
%39$ ms.
(b) Time-evolution of
% instantaneous velocity of the recovery front,
$c_b$ during external stimulation of duration $\tau = 100$ ms
using two different $I$ corresponding to
failed (squares) and successful termination(circles) of activity in the 
medium. 
%The curves are drawn by smoothing the data using a moving window
%average over 5 ms intervals.
During the course of stimulation, $c_f$ is unchanged.
(c) The average waveback velocity $\tilde{c_b}$ (broken lines)
reduces with
%the external sub-threshold current 
$I$ in contrast to $c_f$
(solid lines) 
%[shown for two different durations].
}
\label{Figure3}
\end{figure}

\begin{figure}[t]
\centerline{
\includegraphics[width=1.0\linewidth,clip]{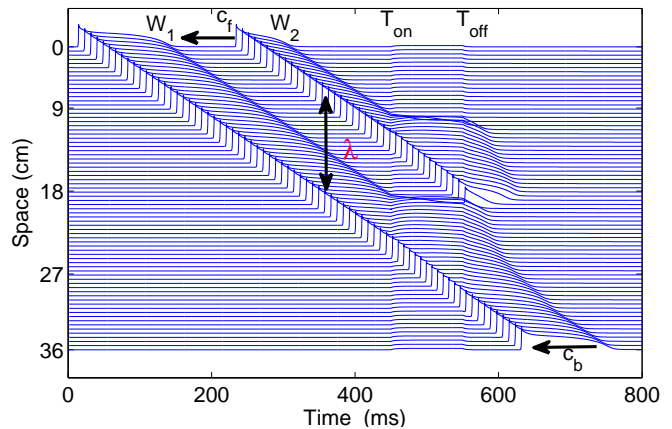} 
}
\caption[]{
%Plot of the membrane potential $V$ showing spatial propagation of two
%waves
Spatiotemporal evolution of two successive waves propagating along 
a 1-dimensional cable of excitable cells when a sub-threshold stimulus
is applied over the time        
interval $[T_{on}, T_{off}]$. 
The propagation
the wave $W_2$ is blocked by the recovery front of $W_1$ due to 
reduction of its velocity, $c_b$. The wavefront velocity $c_f$ of both
waves are almost
unchanged. The spatial interval between the two successive waves at a
time instant is indicated by $\lambda$.
%$I = 1.5 \mu A cm^{-2}, tau = 100 ms$.
}
\label{Figure4}
\end{figure}

In a spatially extended system, this enhanced recovery period of the
cells results in altering the propagation characteristics
%decreasing the waveback velocity 
of the traveling waves. 
Fig.~\ref{Figure3}~(a) shows a spiral wave propagating
in a two-dimensional medium, where each turn of the wave is a region
of excited cells, with the successive turns separated by recovering
regions.
As the state of the cells evolve with time, it is
manifested in space as movement of excitation and
recovery fronts. Their propagation speeds 
are referred to as wavefront ($c_f$) and waveback ($c_b$) velocities,
respectively.
In the absence of any external stimulation, $c_f \simeq c_b$, ensuring
that the width of the excited region remains approximately constant as
the waves travel through the medium.
%(compare the profiles of the
%action potentials in a 1-D cable 
%indicated in Fig.~\ref{Figure4}~(b) by solid and
%dash-dotted lines). 
However, on applying a
sub-threshold stimulus, the waveback velocity becomes significantly
lower than that of the wavefront which is almost unchanged.
Fig.~\ref{Figure3}~(b) shows that, once stimulation begins, $c_b$ quickly
decreases to a minimum value dependent on $I$. It
then gradually rises to eventually become equal to $c_f$ again.
For a large sub-threshold stimulus $I$, the waveback velocity rapidly falls
to its lowest value and changes very slowly thereafter.
Under these conditions, we can ignore the time-variation of
$c_b$ for small $\tau$ and use the time-averaged value
$\tilde{c_b} (I)$.
Increasing $I$ leads to an increased difference in the
velocities of the excitation and recovery fronts, $c_f -
\tilde{c_b} (I)$ (Fig.~\ref{Figure3},~c).
For short stimulus durations, this difference is almost independent
of $\tau$.
%slowing the recovery of cells by using a sub-threshold stimulus
%leads to a gradual increase of the total inexcitable area in the
%The lagging recovery front increases the inexcitable area in the
%medium, thereby making it
%progressively unlikely for the system to sustain recurrent activity.
A significantly lower waveback velocity 
results in the inexcitable region between the excitation and
recovery fronts of a wave becoming extended through the course of the
stimulation (compare the profiles of APs in a 1-D cable shown in
Fig.~\ref{Figure4}). This increases the overall area of the medium
that cannot be excited, thereby making it
progressively unlikely for the system to sustain recurrent activity.

This is explicitly shown for a 1-dimensional cable in
Fig.~\ref{Figure4}. When two successive waves propagate along the
cable, globally applying the sub-threshold stimulus reduces the
excitable gap between the recovery front of the leading wave (whose
velocity $c_b$ has decreased) and the
excitation front of the following wave (whose velocity $c_f$ is
unchanged). 
For a high sub-threshold $I$ applied for a long enough
duration, the waveback of the first wave 
slows down sufficiently to collide with the
succeeding wavefront.
This collision results in termination of the excitation front for the
second wave which subsequently disappears from the medium.
%This picture is fundamentally unchanged with increasing number of
%waves simultaneously existing in the medium. 
%
%Fig.~\ref{Figure4}~(a-b) shows that in a one-dimensional
%cable of coupled excitable cells, the sub-threshold stimulation
%significantly reduces the waveback velocity while the velocity of
%the wavefront is relatively unchanged.
%This increases the inexcitable region in the
%medium by slowing the recovery of the cells. By preventing more and more cells
%from returning to the excitable state by increasing $I$ and $\tau$, we
%cerate a situation where recurrent excitation of the system is made
%increasingly unlikely.
%The main reason is the effect of this on the waveback velocity, while
%wavefront velocity is not affected. 

\begin{figure}[t]
\centerline{
\includegraphics[width=1.0\linewidth,clip]{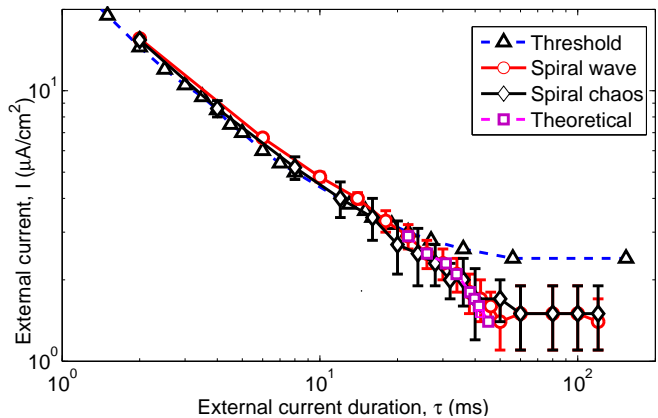}
%Paper_figure5.eps}
}
\caption[]{Strength-duration curves for a two-dimensional medium with the
external current
applied at all points of a quiescent medium (triangles) or a
medium with existing
excitation activity, either a single spiral (circles) or
spatiotemporal chaos (diamond). The theoretical prediction given
in Eq.~(\ref{Eq2}) is also shown (square).
Each ($\tau$,$I$) point is averaged over $10$ initial conditions.
}
\label{Figure5}
\end{figure}

The above physical picture is fundamentally unchanged for a rotating
spiral wave with multiple turns as shown in Fig.~\ref{Figure3}~(a).
We now use this to propose a simple semi-analytical theory 
for the mechanism by which the sub-threshold stimulus
suppresses spatially heterogeneous activity.
%We now use this to propose a simple, semi-analytical theory relating
%the magnitude of the sub-threshold stimulus $I$ and its minimum
%duration $\tau_{min}$ necessary for terminating all activity in the
%medium.
%Let us focus on a particular
%wave, whose excitation front is moving with a propagation velocity
%$c_f$ and the recovering front with a velocity $c_b$.
%In the absence of any external stimulus, $c_f \simeq c_b$.
In the absence of any external stimulus, the width of the excited
region of a wave lying between its excitation and recovery fronts
is
$l = c_f \tau_{r}$, where $\tau_{r}$ is the
period for which the active cells remain excited. 
This time period is operationally measured
as the duration for which the transmembrane potential of a cell ($V$)
remains above its excitation threshold.
%transmembrane potential of a cell, $V > - 60$ mV.
%We also define the length of the excitable gap, i.e., the region
%between the waveback and the front of the succeeding wave, as $l$.
On applying a sub-threshold external current $I$, $c_f$ is almost
unchanged but the resultant waveback velocity, $c_b (I,t)$, 
which varies with time over the duration of the stimulus $\tau$, 
is seen to decrease with increasing $I$.
If $I$ is large or $\tau$ is small, the time-variation
of $c_b$ can be neglected and it is reasonably well-approximated by 
the time-independent average value $\tilde{c_b} (I)$ over the stimulus
duration.
Thus, the width of the excited region of the wave increases
to $l(I) = l + [c_{f} - \tilde{c_b}(I)] \tau$. 
If $\lambda$ is the distance between excitation fronts of two
successive waves in the medium, then collision between the recovery
front of the leading wave and the excitation front of the following
wave takes place when $l(I) \geq \lambda$.
Thus, for a sub-threshold stimulus $I$, the shortest stimulus 
duration $\tau_{min}$ necessary to eliminate a
source of recurrent activity such as a
spiral wave is,
\begin{equation}
\tau_{min} = \frac{\lambda - c_f \tau_r}{c_f - \tilde{c_b}(I)}.
\label{Eq2}
\end{equation}
Eq.~(\ref{Eq2}) provides us with an analytical relation between the stimulus
magnitude and its minimum duration necessary for terminating activity
in the medium in terms of measurable dynamical characteristics of the
system. Fig.~\ref{Figure5} shows that this theoretical
strength-duration curve for the external stimulation necessary to
terminate activity matches very well with the empirical data
obtained from numerical simulations for both single spiral wave as
well as spatiotemporal chaos.
In general, the weaker the sub-threshold current, the longer it has to
be applied in order to alter the dynamical behavior of the system.
However, there is a lower bound for $I$ below which there is no
discernible effect of the sub-threshold stimulation 
regardless of its duration. Note that, for values of $I$ just above
this lower bound, the required $\tau_{min}$ is extremely long and
the temporal variation of $c_b$ over the duration of the stimulation can
no longer be neglected.
By explicitly considering the time-dependence of $c_b$ in
Eq.~(\ref{Eq2}), one can theoretically
estimate the value of $I$ where the strength-duration curve
becomes independent of $\tau$.
%The more exact relationship
%between $l(I)$ and $\lambda$ is obtained by replacing the
%$c_{f}(I)$ and $\tilde{c_b}(I)$ with the instantaneous
%values,so that we have for the condition of annihilation of fronts,
%$\lambda = l + \displaystyle\int_{0}^{\tau_{min}}c_{f}(I,t)\,dt -
%\displaystyle\int_{0}^{\tau_{min}}c_{b}(I,\tau)\,dt$

The mechanism of the sub-threshold response of excitable media
proposed here depends only on the recovery dynamics of the system.
%is robust and only dependent upon the rate of recovery in an excitable
%media model being affected by the external current. In the LR1 model,
In detailed ionic models of biological excitable cells, this
dependence is manifested as a
decrease in the potassium ion channel conductance
responsible for the slow, outward $K^+$ current during the
sub-threshold stimulation.
%this is manifested as a decrease of the potassium ion channel
%conductance responsible for the slow, outward time-dependent $K^+$
%current when the external current is applied.
Thus, simplistic models of excitable media which do not incorporate
the effect of external stimulation on the recovery dynamics are
inadequate to reproduce this 
enhanced sub-threshold response reported here.
%resulting from heterogeneous activity in the medium.
%This is why only seen in models
%where the kinetics reproduces details of the waveback velocity
%dependence on .... (so not seen in FHN models).
Our results provide a framework for explaining earlier experimental
observations that, in the human heart, sub-threshold stimulation not
only prevents subsequent activation~\cite{Prystowsky83}
but also terminates certain types of arrhythmia~\cite{Shenasa88,Salama94}.
The results reported here may have potential significance for
controlling spatiotemporal dynamics in several practical situations
involving excitable media, such as, during clinical treatment of
life-threatening arrhythmias. Current control methods 
primarily use large supra-threshold stimulation to
simultaneously activate all excitable cells. Thus, regions
rendered inexcitable through prior activity are unaffected, and can
subsequently be re-activated by any remaining excitation after the
external stimulation is removed, leading to failure of control.
By contrast, the sub-threshold stimulation method described here
slows the recovery of excited cells, thereby reducing the 
pool of cells available for excitation by existing activity in the
medium.
As this approach, in general,
requires lower energy, 
it suggests a complementary approach for efficiently terminating
spatially extended chaos in excitable systems.
%The proposed mechanism may also provide a key to
%explain how electrical
%intervention can treat life-threatening spatially irregular activity
%in the heart (fibrillation) even though the actual current penetrating
%to the cardiac muscle is extremely low~\cite{Panfilov96,Pumir99}.
The proposed mechanism may also provide a key to understand how
spatially irregular activity in biological systems (e.g.,
fibrillation) can be controlled by signals strongly
attenuated during passage through the intervening
medium~\cite{Panfilov96,Pumir99}.

This work is supported in part by IFCPAR (Project 3404-4) and IMSc Complex
Systems Project (XI Plan).

%{2}

\end{document}